\documentclass[greybox]{svmult}

\usepackage{mathptmx}       
\usepackage{helvet}         
\usepackage{courier}        
\usepackage{type1cm}        
\usepackage{graphicx}        

\begin{document}
\title*{Ferromagnetic models for cooperative behavior: Revisiting {\em Universality} in complex phenomena.}

\author{Elena Agliari\thanks{Dipartimento di Fisica, Universit\`{a} di Parma, \email{elena.agliari@fis.unipr.it}} , Adriano Barra \thanks{Dipartimento di Fisica, Sapienza Universit\`{a} di Roma, \email{adriano.barra@roma1.infn.it}}, Andrea Galluzzi \thanks{Dipartimento di Matematica, Sapienza Universit\`{a} di Roma, \email{galluzzi@mat.uniroma1.it}}, Andrea Pizzoferrato \thanks{Department of Mathematics, University of Warwick, \email{andrea.pizzoferrato@gmail.com}}, Daniele Tantari \thanks{Dipartimento di Matematica, Sapienza Universit\`{a} di Roma, \email{tantari@mat.uniroma1.it}}
}


%

\maketitle

\begin{abstract}
: Ferromagnetic models are {\em harmonic oscillators} in statistical mechanics. Beyond their original scope in tackling phase transition and symmetry breaking in theoretical physics, they are nowadays experiencing a renewal applicative interest as they capture the main features of disparate complex phenomena, whose quantitative investigation in the past were forbidden due to data lacking. After a streamlined introduction to these models, suitably embedded on random graphs, aim of the present paper is to show their importance in a plethora of widespread research fields, so to highlight the unifying framework reached by using statistical mechanics as a tool for their investigation. Specifically we will deal with examples stemmed from sociology, chemistry, cybernetics (electronics) and biology (immunology).
\end{abstract}
%
%

\section{Introduction}

The history of theoretical investigation in collective ferromagnetic behaviors is rooted back in the first decades of the twentieth century, when Lenz introduced a model and -in the winter of 1921- asked to his student Ising to solve it as, for himself, such a research was too trivial. The Second World War, and in particular Nazi persecution, estranged Ising from Germany (and from scientific interchanges) up to late 1947, when, once back, somehow unexpectedly he discovered to be a famous physicist for his contribution in solving the Lenz model (which passed on as the Ising model \cite{barraCW,baxter,thompson}).
\newline
At that time statistical mechanics was developed as a theoretical tool for investigating the structure of matter (solid state physics, liquid and kinetic theories \cite{AM,evans}) and the first concept of ``universality"  \cite{ellis} highlighted how different {\em physical} systems behave in a very similar way close to criticality. Other decades had to elapse before the scientific community started to realize that such "universal" behavior was far from being restricted to the physical scenario, and a mature understanding that "several element showing imitative interactions" may behave collectively as a ferromagnet -whatever the context- is nowadays achieved.
\newline
However, the boundaries of validity of the last assertion are still under investigation as our knowledge of ferromagnetism is growing and merging with "imitation" \cite{daurlauf},  "cooperation" \cite{chemkin}, "amplification" \cite{millmann}, "syncronization" \cite{kuramoto}, etc. Moreover, disparate fields of sciences continually spring up: lapping (a part of) such boundaries is the focus of the present paper.
\newline
What we need, as a theoretical benchmark, is the description of an ensemble of dichotomic spins living on the nodes of random graphs. Hence, we first provide a streamlined introduction to the statistical mechanics of ferromagnetism and a minimal smattering regarding the underlying graph theory. Then, we turn to extrapolate such an imitative behavior from the real world, stemming examples from several fields of science as sociology, chemistry, cybernetics (electronics) and biology (immunology). Summarizing, we are going to show that

\begin{itemize}
\item in sociology, focusing on a test-case among many \cite{daurlauf, BC1}, namely the phenomenon of social integration of migrants inside a host community, we are going to analyze as a standard quantifier the amount of mixed marriages (where ``mixed" means achieved by a native and a migrant): we will show that, once plotted against the percentage of migrants inside the host community, its behavior is 
identical to the one of observable typical of statistical mechanics (i.e. the magnetization versus the temperature), 
highlighting the key role -in imitative behavior- played by each agent belonging to the community. We will show how (and why) this phenomenon can be reabsorbed within the ferromagnetic phenomenology \cite{pnas}.

\item in chemistry, in particular dealing with reaction kinetics as a concrete example, many polymers and proteins exhibit cooperativity, meaning that their ligands bind in a non-independent way: if, upon a ligand binding, the probability of further binding (by other ligands of the same protein/polymer) is enhanced, like in the paradigmatic case of hemoglobin \cite{thompson}, the cooperativity is said to be positive. As we are going to show, such a cooperative behavior in chemical reactions can be perfectly described by the statistical mechanics of ferromagnetism \cite{ScRep2,noi_cin_kim}.

\item in electronics, we find the hallmark of ferromagnetic behavior already in its fundamental bricks, namely in operational amplifiers \cite{millmann}. As we will show, there is a one-to-one mapping between self-consistency in statistical mechanics and transfer function in electronics. In particular, when no amplification is present, such a transistor can be mapped into an ensemble of non-interacting spins, but, when the circuit is amplifying the input signal (hence the output is proportional to the input by a constant of proportionality larger than one) interaction among its constituents can be mapped into interaction among spins and, again, its behavior is perfectly described by means of the statistical mechanics of ferromagnetism \cite{ScRep2}.

\item in immunology, B clones (namely the ensemble of identical B cells producing the same antibodies) can interact reciprocally by imitation (that immunologists call ``elicitation"): if  a clone undergoes expansion and antibody release, its nearest neighbor (in the idiotypic network, namely the random graph whose nodes are the B clones and whose links are their reciprocal strengths of interaction \cite{janaway,noi_JSTAT}) will also undergo clonal expansion and antibody release too. Again, such a behavior is remarkably captured by the statistical mechanics of ferromagnetism \cite{anergy}.
\end{itemize}

Before proceeding, we notice that the three-dimensional Ising model is still unsolved, and enormous efforts have been necessary, e.g. by Onsager and followers \cite{onsager, baxter}, in order to solve the model at low dimensionality. However, for all our examples, and away from the physical world (where the power-laws  
of gravity and electromagnetic fields strongly require projection on two- and three-dimensional structures), we will deal with the so called "mean-field" approximation. The latter is completely solvable as it assumes spins interacting broadly on random graphs (e.g. Erd\"{o}s-R\'{e}nyi topologies \cite{randomgraphs}) instead of peer-to-peer physical interactions on lattices: while this feature constitutes an approximation in the pure-physical community, in all the branches of science we outline (as well as in several others), where interactions are not short-ranged, this is perfectly reasonable, at both theoretical and empirical levels. Indeed, the mean-field statistical mechanics, revealed itself as a powerful and unifying instrument to investigate the complexity of our world: our understanding of collective behaviors by interacting agents from this perspective is an extremely exciting research field,  still at the beginning, and we believe statistical mechanics will become a stronger and stronger technology for this task in the near future.
%
%
\section{Definition of the model and thermodynamics}

Let us consider an ensemble of $N$ agents (spins), whose state is represented by a dichotomic variable $\sigma_i = \pm 1$, with $i \in (1,...,N)$; through the paper, spins will assume a different meaning according to the context. Agents interact with each other, if reciprocally connected, via a positive coupling $J$, hence we can write an Hamiltonian for the system as
\begin{equation}
H_N[\sigma|J]=-\frac{1}{N}\sum_{i<j} J_{ij} \sigma_{i}\sigma_{j} - h \sum_i \sigma_i,
\end{equation}
where $h$ is an external scalar field (magnetic in the physical literature) and the coupling $J_{ij}$ is set equal to either $1$ or $0$ according to a given probability distribution. This choice automatically frames the model on an Erd\"{o}s-R\'{e}nyi graph \cite{smallworld}.
By imposing $J_{ij}=1$, when the link between $i$ and $j$ exists, we only lock the temperature scale without changing the physics of the problem. Of course $J_{ij}=0$  simply means that the two corresponding nodes are not interacting.
\newline
The role of dilution, from a statistical mechanics perspective, at least at the level of the mean values of observable, is simply to reduce the averaged strength reciprocally felt by the spins, but does not alter\footnote{as far as 
the network remains over-percolated. If the percolation threshold is crossed, the system splits into independent subsystems
and the analysis reduces to the sum of the analysis on each subsystem.} the physics \cite{guerra2,SM_Jstat}.
\newline
The thermodynamic of the model is carried by the free energy density $\lim_{N \to \infty}f_N(\beta)=\lim_{N \to \infty} F_N(\beta)/N$ , which is related to the Hamiltonian via
\begin{equation}
e^{-\beta F_N(\beta,h)}=Z_N(\beta,h)=\sum_{\{\sigma\}} e^{\frac{\beta}{N}\sum_{i<j}\sigma_{i}\sigma_{j}+ \beta h  \sum_i \sigma_i},
\end{equation}
where $Z_N(\beta,h)$ is the partition function. For the sake of convenience we will
not deal with $f_N(\beta,h)$ but with the thermodynamic pressure $A(\beta,h)$ defined via
\begin{equation}
A(\beta,h)= \lim_{N\to\infty}A_N=\lim_{N \to \infty} -\beta f_N(\beta,h)=\lim_{N \to \infty}\frac{1}{N}\ln Z_N(\beta,h).
\end{equation}
A key role will be played by the order parameter, namely the magnetization $m$, that reads as
\begin{equation}
m_N=\frac{1}{N}\sum_{i=1}^N \sigma_i,\qquad \langle m_N\rangle=\frac{\sum_{\{\sigma\}} m_N e^{-\beta H_N[\sigma]}}{\sum_{\{\sigma\}} e^{-\beta H_N[\sigma]}},
\end{equation}
where in the last definition the brackets $\langle . \rangle$ denote the Boltzmann average.
\newline
Note that the order parameter, namely a single function of the tunable parameters that describes the ``typical"
behavior of the system, is nothing but the arithmetic average of all the single degrees of freedom the system may use to
respect thermodynamics.

\begin{figure}[h!!]
\sidecaption
\includegraphics[scale=.6]{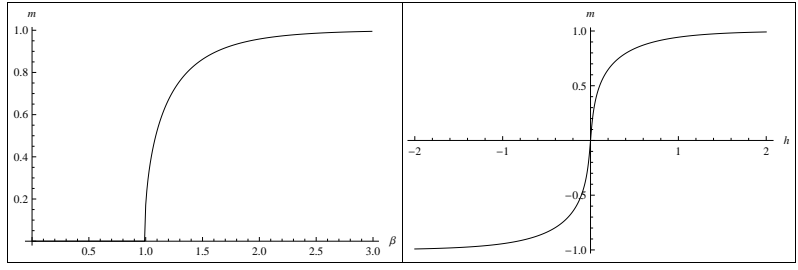}
\caption{Left panel: the behavior of the order parameter $m$ as a function of the noise level $\beta$ can be formally described by a square-root function. Right panel: the behavior of the order parameter $m$ as a function of the external field $h$ can be formally described by a hyperbolic tangent, i.e. a sigmoidal curve.}\label{fig:Ising}
\end{figure}

In order to solve for the free energy (strictly speaking for the pressure), namely to obtain an explicit functional expression of $A(\beta,h)$ in terms of the tunable parameters $\beta$ and $h$ and of the order parameter $m$, we are going to use Guerra's interpolation scheme \cite{barraCW,guerra1,guerra2}. The idea behind this approach is to interpolate between the original system and a system of independent spins interacting with an effective field able to simulate fictitiously the stimuli induced by the others. To this task we introduce the following interpolating Hamiltonian
\begin{equation}
H(t)=t H_{original} + \left(1-t \right) H_{one-body},
\end{equation}
where $t \in [0,1]$ is the interpolation parameter, and the corresponding (time dependent) partition function $Z_N(t)$, pressure $A_N(t)$ and Boltzmann state $\langle m_N\rangle_t=Z_N^{-1}(t)\sum_{\{\sigma\}} m_N e^{-\beta H(t)}$. Choosing $H_{one-body}[\sigma]=-\bar{m}\sum_i\sigma_i$, once introduced a trial parameter $\bar{m}$ to be optimized at the end of the procedure, we can use the fundamental theorem of calculus applied to the pressure:
\begin{equation}\label{eq:AdA}
A_N=A_N(1)=A_N\left(0\right)+\int_{0}^{1}\frac{d A_N(t)}{d t}dt,
\end{equation}
By a direct calculation is then trivial to show that the pressure of the ferromagnetic model in Guerra's interpolation scheme is given by
\begin{eqnarray}
A_N&=&\ln2+\ln\left[\cosh\left(\beta\bar{m}\right)\right]-\left(\frac{\beta}{2}\right)\bar{m}^{2}+\int_{0}^{1}\left(\frac{\beta}{2}\right)\left<\left(m_N-\bar{m}\right)^{2}\right>_t dt\nonumber \\
&=&A^{trial}(\bar{m})+\int_{0}^{1}R_N(t;\bar{m}) dt,
\end{eqnarray}
where $A^{trial}(\bar{m})=\ln 2+\ln\left[\cosh\left(\beta\bar{m}\right)\right]-\left(\frac{\beta}{2}\right)\bar{m}^{2}$ .
The rest  $R_N(t;\bar{m})=\beta/2 \langle (m_N - \bar{m})^2 \rangle_t$, that one wants to remove or reduce as possible,  is positive defined and represents the fluctuation of the magnetization around $\bar{m}$. Since in the thermodynamic limit the magnetization is a self averaging order parameter, it is possible to find an optimum $\bar{m}$ such that  $P(m)=\delta(m-\bar{m})$ and consequently $\lim_{N\to\infty}R_N(t;\bar{m})=0$.  From the positivity of the rest, it is easy to see that the optimum $\bar{m}$ can be found by minimizing the trial free energy $-\beta^{-1} A^{trial}(\bar{m})$. In this way we obtain the self-consistent equation which rules the behavior of the order parameter itself (from the previous considerations we can argue that the optimal trial parameter $\bar{m}$ assumes the physical meaning of the thermodynamic limit of the system's magnetization itself, i.e. $\bar{m}=\lim_{N\to\infty}\langle m_N(\sigma)\rangle:=\langle m\rangle$):
\begin{equation}
\langle m\rangle =\tanh\left[\beta(\langle m\rangle+h)\right].
\end{equation}
In order to simplify the understanding of the bridges we pursue, it is convenient to plot the behavior
of the order parameter versus the two tunable parameters, noise level $\beta$ and external field $h$ (see fig.~\ref{fig:Ising}).

%
%
\newpage
%
%
\section{Ferromagnetic behavior in quantitative sociology}

\begin{figure}[h!!]
\sidecaption
\includegraphics[scale=.55]{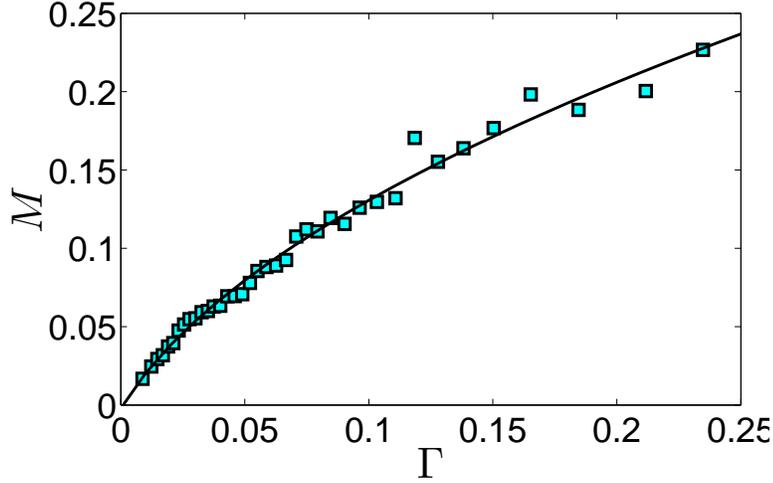}
\caption{The plot shows the (normalized) amount of mixed marriages versus the percentage of migrants
inside the host country: squares are real data mirroring the decade $1990-2000$ in Spain while the continuous black curve
is the best fit with a trial $f(\Gamma)= a \sqrt{\Gamma}$ with $a\sim0.52\pm0.02$ and an $R^2 \sim 0.992$.
Note that here $m$ plays as the magnetization in the mean-field Ising model while $\Gamma$ plays as the (rescaled) temperature.
Data are extracted from \cite{pnas}.}\label{fig:matrimoni}
\end{figure}

In the following we briefly summarize results obtained in the analysis of social networks, particularly focusing (for the sake of concreteness) on immigration phenomena \cite{BC1}, reporting a quantitative result from \cite{pnas}.
\newline
In this context we want to show that classical integration quantifiers like the percentage of mixed marriages,
once plotted versus the percentage of migrants inside the host community,
behaves as the magnetization versus the temperature of classical mean-field ferromagnetism, namely with the order parameter scaling as a square root of the tunable noise level.
Calling $N_{imm}$ the amount of immigrants in the host country and the total population (of immigrants and natives) $N = N_{imm} + N_{nat}$ and defining $\gamma = N_{imm}/N$, a natural parameter for assessing change in integration quantifier is the product $\Gamma \equiv \gamma(1-\gamma)$ as
\begin{equation}
N_{imm} N_{nat} \propto \Gamma,
\end{equation}
since it counts the number of possible cross-group links. By analyzing a database on immigration and integration from
Spain in the time window $1990-2000$, we found that the quantifier capturing the mixed-marriages displays non-linear behavior, in particular it follows remarkably a square root (see fig.~\ref{fig:matrimoni}).

Understanding such a behavior from a statistical mechanics perspective is quite simple:
let us consider two ensembles of agents, the natives, denoted by $\pm 1 \ni \sigma_i$, $i \in (1,...,N)$ and the immigrants, denoted by $\pm 1 \ni \tau_{\mu}$, $\mu \in (1,...,P)$. Of course $\gamma = P/(P+N)$. The values $\pm 1$ coupled to the possible values of $\sigma$ and $\tau$ stand for a positive attitude (+1), or its lack (-1), with respect to contracting a marriage with an immigrant and a native, respectively.
If we believe that imitation plays a role in social networks, it is then possible to built an Hamiltonian as
\begin{equation}
H(\sigma,\tau) = \frac{-1}{P+N}\sum_{i,\mu}^{N,P}\sigma_i \tau_{\mu},
\end{equation}
that represents the following: stable (potential) couples are those where the members are both happy or unhappy with the mixed marriage. What is unfavorable is a long-term state where one of the two members wants the mixed marriage but the other does not.
We can then built the statistical mechanics machinery to see what this prescription implies. The partition function reads off as
\begin{equation}
Z(\beta, \gamma) = \sum_{\sigma}\sum_{\tau} \exp\left( \frac{1}{P+N}\sum_{i,\mu}^{N,P}\sigma_i \tau_{\mu} \right) \sim \sum_{\sigma} \exp\left( \frac{\gamma(1-\gamma)}{2N}\sum_{ij} \sigma_i \sigma_j \right),
\end{equation}
which is nothing but the partition function of a (single party) ferromagnetic model with coupling $J = \gamma(1-\gamma)$.
\newline
Following the previous section (as we reduced to that framework) we know how to write the self-consistency, which reads here as
\begin{equation}
\langle m \rangle = \tanh\left[ \beta \gamma(1-\gamma) \langle m \rangle \right] \sim \beta\sqrt{\gamma(1-\gamma)}.
\end{equation}
Hence, if imitation has a key role in social networks we expect that the average attitude of the population versus the percentage of migrants, and, ultimately the number $M$ of mixed marriages, depends on $\Gamma$ as $M \sim \sqrt{\Gamma}$, exactly as we experimentally found in this test-case (see fig.~\ref{fig:matrimoni}).
%
%
%
%

\section{Ferromagnetic behaviors in biochemistry}

Chemical kinetics usually considers a hosting molecule $P$ that can bind $N$ identical molecules $S$ on
its structure; calling $P_j$ the complex of a molecule $P$ with $j \in [0:N]$ molecules attached, the reactions leading to the
chemical equilibrium are the following: $S+P_{j-1}\rightleftharpoons P_j$, and, as a convenient experimental observable, usually the average number $\overline{S}$ of substrates bound to the protein is considered as
\begin{equation}\label{eq:adair}
\overline{S}=\frac{\sum_{i=1}^N i [P_i]}{\sum_{i=1}^N [P_i]}=\frac{\sum_{i=1}^N i K^{(i)}[S]^i}{1+K^{(1)}\sum_{i=1}^N K^{(i)}[S]^i},
\end{equation}
which is the well-known Adair equation \cite{ScRep2}, whose normalized expression defines the $\emph{saturation function}$ $Y = \overline{S}/N$.
More generally, one can allow for a degree of sequentiality and write
\begin{equation}
 \overline{Y}=\frac{K[S]^{n_H}}{1+K[S]^{n_H}},
\end{equation}
which is the well-known Hill equation \cite{noi_cin_kim}, where $n_H$, referred to as Hill coefficient, represents the effective number
of substrates which are interacting, such that for $n_H=1$ the system is said to be \emph{non-cooperative} and the
Michaelis-Menten law is recovered while for $n_H>1$ it is \emph{cooperative}.
In order to bridge this scenario with statistical mechanics, following \cite{ScRep2}, let us consider an ensemble of elements (e.g. identical macromolecules, homo-allosteric enzymes, etc.),
whose interacting sites are overall $N$ and labeled as $i = 1,2,...,N$. Each site can bind one smaller molecule (e.g. of
a substrate) and we call $\alpha$ the concentration of the free molecules ($[S]$ in standard chemical kinetics language as used before). We
associate to each site an Ising spin such that when the $i^{th}$ site is occupied $\sigma_i = +1$, while when it is empty $\sigma_i = -1$.
A configuration of the elements is then specified by the set $\{\sigma\}$.

\begin{figure}[h!!]
\sidecaption
\includegraphics[scale=.55]{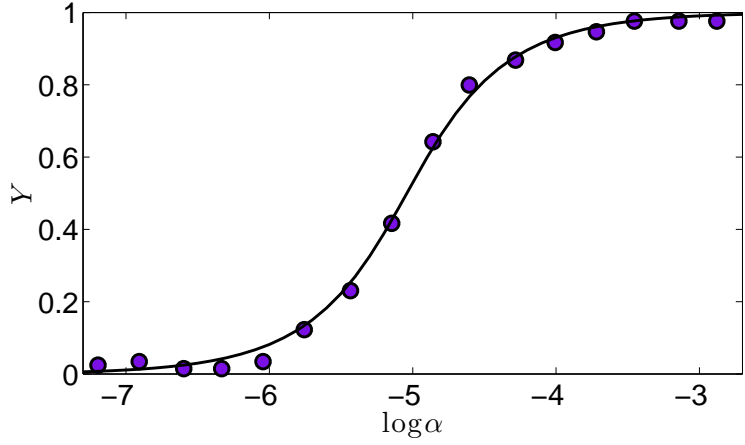}
\caption{The plot shows the saturation curve (the  normalized amount of bind ligands) versus the log-concentration
of the substrate in a simple titration. Real data ($\circ$) representing $\textrm{Ca}^{2+}$-calmodulin-dependent protein kinase are best fitted (solid line) using the self-consistency $1.16$, with $R^2 \sim 0.98$.
Note that here $Y$ plays as the (shifted) magnetization of the mean-field Ising model, while $\log(\alpha)$ plays
as the external field $h$. Data are extrapolated from \cite{ScRep2}.}\label{fig:chimica}
\end{figure}

We model the interaction between the substrate and the binding site by an external field $h$ meant as a proper measure for the concentration of free-ligand molecules, hence $h = h(\alpha)$.
We can think at $h$ as the chemical potential for the binding of substrate molecules on sites: when it is positive,
molecules tend to bind to diminish energy, while when it is negative, bound molecules tend to leave occupied sites. The chemical potential can be expressed as the logarithm of the concentration of binding molecules and one can assume
that the concentration is proportional to the ratio of the probability of having a site occupied with respect to that
of having it empty, and we can pose $h=\frac{1}{2}\ln(\alpha)$.

Similarly to the previous test-case drawn from sociology, here we focus again on pairwise interactions and we use complete bipartite graph structure. Sites are divided in two groups, referred to as $A$ and $B$, whose sizes are $N_A$ and $N_B$ $(N = N_A + N_B)$, respectively.
Each site in $A$ $(B)$ is linked to all sites in $B$ $(A)$, but no link within the same group is present.
As a result, given the parameter $J$ and $h$, the energy associated to the configuration ${\mathbf{\sigma}}$ turns out be
\begin{equation}\label{eq:Hck}
H_N[\{\mathbf{\sigma}\};J,h]=-\frac{1}{N_A+N_B}\sum_{i=1}^{N_A}
\sum_{j=1}^{N_B}J \sigma_{i}\sigma_{j}-h\sum_{i=1}^{N_A+N_B}\sigma_i.
\end{equation}
Note that in (\ref{eq:Hck}) the sums run over all the binding sites: despite we deal with the thermodynamic limit, this does not imply that we model macromolecules of infinite length,
which is somehow unrealistic, but that we can consider $N$ as the total number of binding sites, localized even on different macromolecules, as boundary effects can be reabsorbed in an effective renormalization of the couplings $J \geq 0$.

A key point is that the saturation function $Y(\alpha)$ is closely related to the magnetization in statistical mechanics $m(h)$
as it reads off as
$$
Y(\alpha)=\frac12 \sum_{i=1}^N (1 + \sigma_i) = \frac12 [1+\langle m (h(\alpha))\rangle].
$$
Recalling the expression for the self-consistency equation, we are immediately able to  see that $Y(\alpha)$ fulfills the following free-energy minimum condition
\begin{equation}\label{eq:Y}
Y(\alpha)=\frac{1}{2}\{ 1+ \tanh[J(2Y-1)+  \frac{1}{2}\log(\alpha)] \}.
\end{equation}
This expression returns the average fraction of occupied sites corresponding to the equilibrium state for the system.
In general, the Hill coefficient can be obtained as the slope of $Y (\alpha)$ in eq. (\ref{eq:Y}) at the symmetric point $Y = 1/2$,
namely
\begin{equation}\label{eq:nH}
n_H=\frac{1}{Y(1-Y)}\frac{\partial Y}{\partial \alpha}\bigg|_{Y=1/2}=\frac{1}{1-J}.
\end{equation}
Further, the expression in eq. (\ref{eq:Y}) can be used to fit experimental data for saturation versus substrate concentration.
As shown in fig.~\ref{fig:chimica}, the fit of experimental data is very good and Hill
coefficients derived in this way and the related estimates found in the literature are also in excellent agreement.
%
%
%
%

\section{Ferromagnetic behaviors in electronics}

 This section is dedicated to the understanding, within the statistical mechanics framework, of collective behaviors in
 cybernetics; in particular, we focus on the electronic declination of cybernetics because this is probably the most
 practical and known branch \cite{millmann}. 

Following \cite{ScRep2}, the plan is to compare self-consistencies in statistical mechanics and transfer functions in electronics so to reach
a unified description for these systems.

The core of electronics is the operational amplifier, namely a solid-state integrated circuit (transistor), which uses feed-back regulation to set its functions.
\newline
An ideal amplifier is the \emph{linear} approximation of the saturable one and essentially assumes that the voltage at
the input collectors is always at the same value so that no current flows inside the transistor \cite{millmann}.

\begin{figure}[h!!]
\sidecaption
\includegraphics[scale=.45]{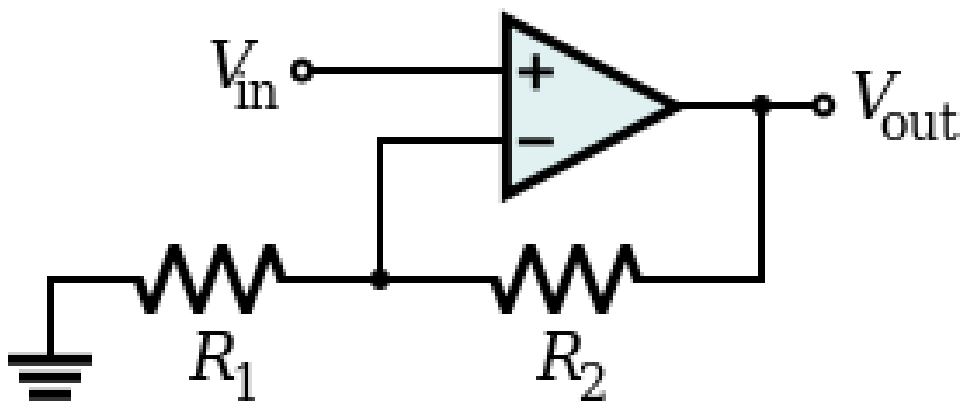}\includegraphics[scale=.25]{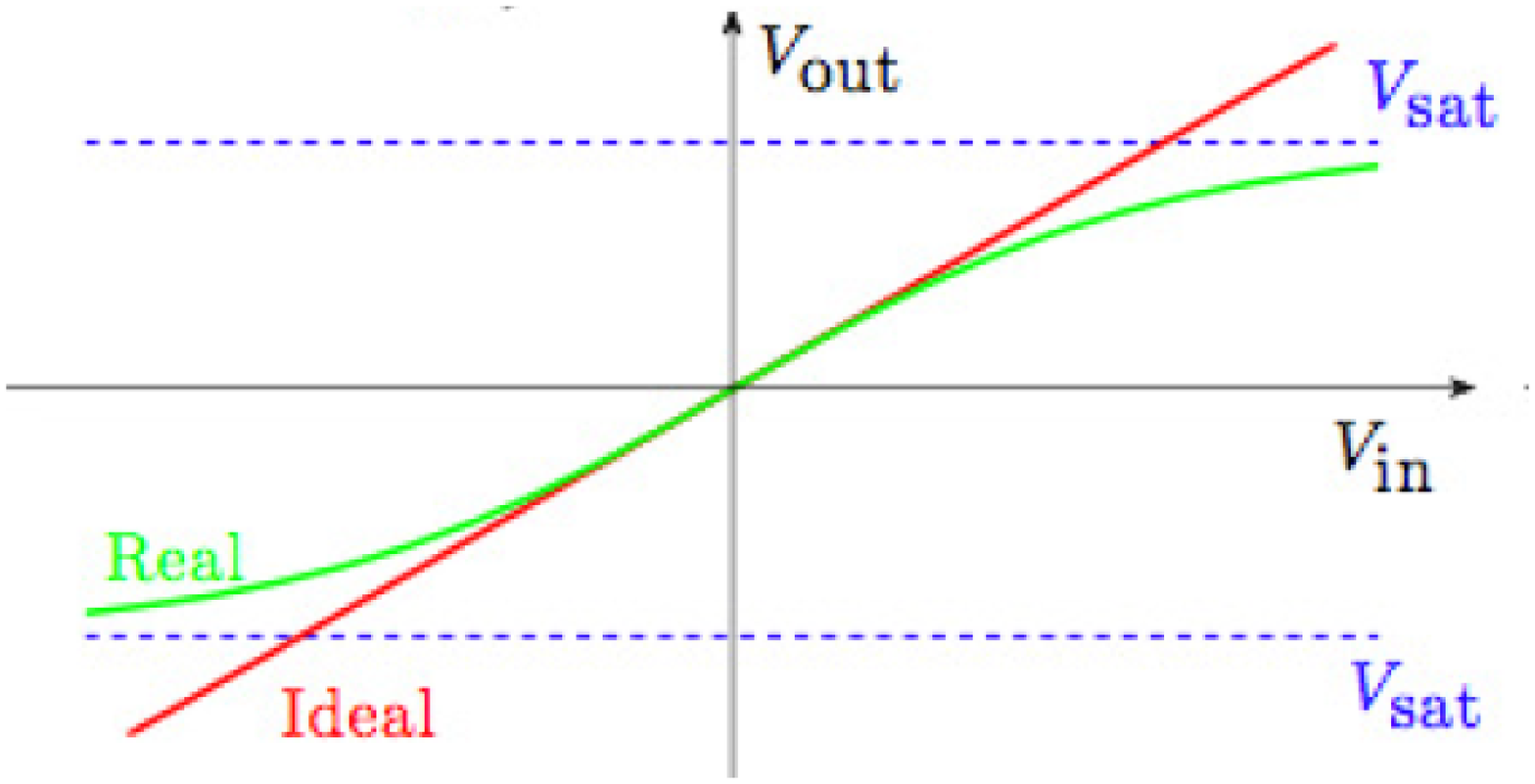}
\caption{Left panel: Schematic representation of an operational amplifier. Note that $R_2$ modulates the strength of the
feed-back signal, the retroaction that allows amplification, as $J$ modulates the feed-back of a spin via its nearest
neighbors in statistical mechanics. Right panel: Transfer function, namely $V_{out}(V_{in})$. Red line is the `ideal response",
where the transistor linearly amplifies, while the green curve represents the real transfer function, showing the
collapse to the asymptotes $\pm V_{sat}$, where $V_{sat}$ is the saturation value reached by the amplifier. Note the manifest
behavior a' la hyperbolic tangent, typical of ferromagnetism.}\label{fig:cibernetica}
\end{figure}

If we call $V_{out}$ the output signal (in Volts) and $V_{in}$ the input signal (in Volts) that exits/enters the
amplifier, and we call $R_2$ (in $\Omega)$ the resistor that allows for retroaction
(feed back signal), then the transfer function  of the system
can be obtained as $V_{out}=(1+R_2)V_{in}$ \cite{millmann} (without loss of generality we set $R_1= 1 \Omega$ for the external
resistor, see fig.~\ref{fig:cibernetica} (left)).
Therefore, as far as $R_2 > 0$, the gain is larger than one and the circuit is amplifying the input.

To highlight our parallel, we note that the transfer function is an input/output relation, exactly as the equation for
 the order parameter $m$. In fact, for small values of the coupling $J$ (so to mirror ideal amplifier), we can write
\begin{equation}
\langle m\rangle\sim(1+J)h.
\end{equation}
Thus, the external signal $V_{in}$ is replaced by the external field $h$, and the response of the system $V_{out}$ is
replaced by the response of the system $\langle m\rangle$.
By comparison we see that $R_2$ plays as $J$, and, consistently, if on the
electronic side $R_2 = 0$ the retroaction is lost and the gain is no longer possible: This is perfectly consistent
with the statistical mechanics perspective, where if $J = 0$ spins do not mutually interact and no feed-back is allowed.
The sigmoidal shape of the hyperbolic tangent is not accounted by ideal amplifiers: this is because saturation is
not included in the approximation we discussed, however, it simply makes $V_{sat}$ asymptotes for the growth, hence
recovering the expected behavior, as shown in fig.~\ref{fig:cibernetica} (right).
%
%
%
%

\section{Ferromagnetic behaviors in theoretical immunology}

Concerning cooperation in biology, we focus on the field of immunology, as we spent some
years studying the emerging collective behavior of lymphocytes, see e.g. \cite{JTB,PRL1}.
\newline
The immune system is a marvelous and extremely complex ensemble of different cells and
signalling proteins: we will focus only on a sub-shell of the whole system, namely the population of B-cells, the soldiers dedicated to the antibody production. Classical clonal selection theory \cite{janaway} assumes that  a host-body has an enormous amount of different B-cells producing different antibodies. B-cells producing the same antibody are grouped into "clones" and the collection of all the clones forms the ``repertoire". Clones have the peculiarity, beyond antigenic recognition, to respond also to stimulation from other lymphocytes: if a clone is releasing antibodies and those are complementary enough to the receptors of another clone, the latter will start to release antibodies as well. This mechanism, called ``elicitation" in immunology, strongly resembles imitative behavior and gives rise to the so called Jerne idiotypic network \cite{noi_JSTAT}, whose properties we want briefly to outline.

\begin{figure}[h!!]
\sidecaption
\includegraphics[scale=.55]{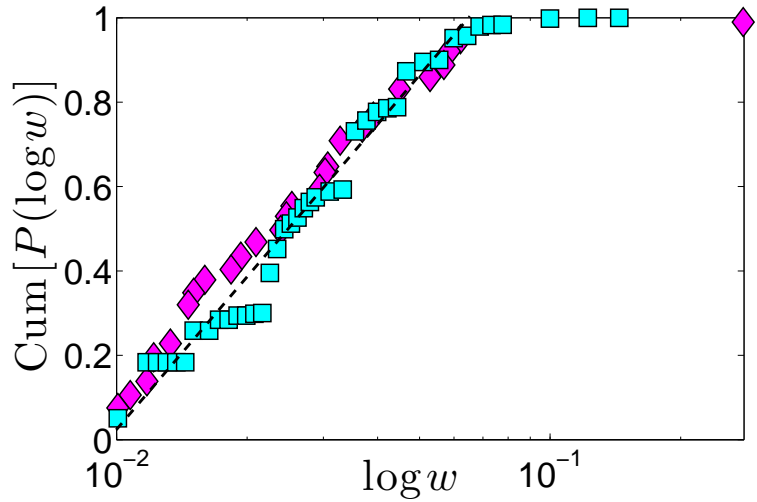}
\caption{Plot shows the cumulative distribution (frequency for the experimental part)
of the weighted connectivity shown by B-lymphocytes in mice. Data are elaborated from \cite{anergy}: Rumbles are
experimental data, squares are Monte Carlo simulations with our $P(W)$ and the dashed black line represents the
theoretical scaling of $P(W)$.}\label{fig:carneiro}
\end{figure}

Proceeding along a general information theory perspective, we associate to each antibody, labeled as $i$,
a binary string $\Psi_i$ of length $L$, which effectively carries information on its structure and on its ability to form complexes with other antibodies or antigens. Since antibodies secreted by cells belonging to the same clone share
the same structure, the same string $\Psi_i$ is used to encode the specificity of the whole related B clone.
In this way, the repertoire will be represented by the set $N$ of properly generated
strings.
Antibodies can bind each-other through ``lock-and-key" interactions, that is, interactions are mainly hydrophobic and electrostatic and chemical affinities range
over several orders of magnitude \cite{janaway}. This suggests that the more complementary two structures are and the more
likely (on an exponential scale) their binding. We therefore define $\chi$ as a Hamming distance,
to measure the complementarity between two bit-strings and introduce a phenomenological coupling
\begin{equation}\label{logj}
\chi_{ij}=\sum_{k=1}^L [\Psi_i^k (1-\Psi_j^k) + \Psi_i^k (1-\Psi_j^k) ] \Rightarrow J_{ij}\propto e^{\alpha \chi_{ij} }
\end{equation}
where $\alpha$ tunes the interaction strength.
\newline
Hence, different clones interact with external antigens and among each other with a coupling given by their reciprocal binding affinities of the corresponding antibodies.
The latter can be formalized in Hamiltonian terms as follows:
\begin{equation}
H_N(\sigma|J) = - \frac{1}{N}\sum_{i<j}J_{ij}\sigma_{i} \sigma_{j} - \sum_{i} h_{i} \sigma_{i},
\end{equation}
where $N$ is the total number of different clones (the size of the repertoire), the dichotomic spin $\sigma_i$ may assume values $+1$
representing antibody release or $-1$ representing quiescence and the positive coupling $J_{ij} \geq 0$ ensures
reciprocal elicitation (imitation) when different from zero, and $h_i$ represents the antigenic load (implying a response by node $i$).
\newline
Now a crucial observable is the weighted connectivity, defined as $W_i = \sum_j^N J_{ij}$, whose distribution $P(W)$,
exploiting the fact that couplings are log-normally distributed (see eq. ($\ref{logj}$) and \cite{noi_JSTAT,anergy}),
can be approximated as
\begin{equation}
P(W)\sim \frac{1}{W \tilde{\sigma} \sqrt{2\pi}}e^{-\frac{(\ln(W)-\tilde{\mu})^2}{2\tilde{\sigma}^2}},
\end{equation}
where $\tilde{\mu}$ and $\tilde{\sigma}$ are related to the $J_{ij}$ distribution
(a detailed derivation of these values can be found in \cite{anergy}).
The last observable deserves attention as it can be compared with experimental results performed on
ELISA technology on mice, as reported in fig.~\ref{fig:carneiro}, depicting data from \cite{anergy}: again, there is a remarkable
agreement between real data and ferromagnetic predictions.

\section{Summary}

In view of broad applications, suitable to help the scientific community in properly framing complexity of Planet Earth into major scaffolds, in these notes we revised the paradigmatic ferromagnetic mean-field scenario, embedding positive coupling among spins on a random graph, such that nodes represent spins and links, whenever present, mirror their interactions.
\newline
Beyond a classical role in depicting the essence of phase transitions and spontaneous symmetry breaking in theoretical physics, this model, and more properly the statistical mechanics approach to model cooperativity, is finding a renewed role in tackling the emergent behavior of disparate systems as a function of tunable external parameters. Indeed, applications have focused on a broad range of systems, all sharing the same microscopic structure, made of by several interacting elements (theoretically denoted as ``spins" and whose nature is specified by the particular example considered) via positive (imitative) couplings.

In these notes we showed, trough several examples and comparisons with real data, that in chemistry (with the example of reaction kinetics, where spins are ligands), in biology (stemming from the idiotypic network of lymphocytes where spins are B-clones), in sociology (by investigating migrant's integration inside a host community where spins are the decision makers) and in electronics (analyzing the transfer function of operational amplifiers, where spins are internal junctions), the statistical mechanics of ferromagnetism is able to properly describe the complex, emergent, phenomenology of their order parameters.
\newline
Hence, the role of this review is to highlight a key, unifying, role performed by this technique in showing that systems apparently diverse and unrelated, behave in the same way once properly described. We believe that merging separate disciplines by finding a ``universal behavior" is an important requisite in order to quantify the complexity of such fields, which, ultimately, reflects the complexity of Planet Earth, focus of the present volume.

\section*{Acknowledgements}

The authors acknowledge the FIRB grant RBFR08EKEV, Sapienza Universit\'{a} di Roma, INFN and INdAM for financial support.
\newline
Our colleagues, alphabetically ordered as Raffaella Burioni, Pierluigi Contucci, Gino Del Ferraro, Aldo Di Biasio, Francesco Guerra, Francesco Moauro, Richard Sandell, Guido Uguzzoni and Cecilia Vernia, are truly acknowledged for walking together into this fascinating research route.

\end{document}